\documentclass[aps,prl,reprint, superscriptaddress]{revtex4-1}
\usepackage{blindtext}
\usepackage{graphics}
\usepackage{graphicx}
\usepackage{amssymb}
\usepackage{amsmath}

\usepackage{caption}

\usepackage{array}
\newcolumntype{L}[1]{>{\raggedright\let\newline\\\arraybackslash\hspace{0pt}}m{#1}}
\newcolumntype{C}[1]{>{\centering\let\newline\\\arraybackslash\hspace{0pt}}m{#1}}
\newcolumntype{R}[1]{>{\raggedleft\let\newline\\\arraybackslash\hspace{0pt}}m{#1}}

\begin{document}

\title{Insulating Josephson-junction chains as pinned Luttinger liquids}
	
\author{Karin Cedergren}
	\affiliation{Centre for Engineered Quantum Systems (EQuS), School of Physics, University of New South Wales, Sydney 2052, Australia}
\author{Roger Ackroyd} 
	\affiliation{Centre for Engineered Quantum Systems (EQuS), School of Physics, University of New South Wales, Sydney 2052, Australia}
\author{Sergey Kafanov}
\altaffiliation[Present address: ]
{Physics Department, Lancaster University, Lancaster, UK \hspace{2pt} LA1 4YB.} 
	\affiliation{Centre for Engineered Quantum Systems (EQuS), School of Physics, University of New South Wales, Sydney 2052, Australia}

\author{Nicolas Vogt}
	\affiliation{Chemical and Quantum Physics, School of Science, RMIT University, Melbourne 3001,VIC 3001 Australia}
\author{Alexander Shnirman} 
	\affiliation{Institut f\"ur Theorie der Kondensierten Materie, Karlsruhe Institute of Technology, D-76128 Karlsruhe, Germany}
	\affiliation{Landau Institute for Theoretical Physics, 119334 Moscow, Russia}
\author{Timothy Duty}
\email[To whom correspondence should be addressed; E-mail: ]{t.duty@unsw.edu.au}
	\affiliation{Centre for Engineered Quantum Systems (EQuS), School of Physics, University of New South Wales, Sydney 2052, Australia}

\begin{abstract}
Quantum physics in one spatial dimension is remarkably rich, yet even with strong interactions and disorder, surprisingly tractable. This is due to the fact that the low-energy physics of nearly all one-dimensional systems can be cast in terms of the Luttinger liquid, a key concept that parallels that of the Fermi liquid in higher dimensions. Although there have been many theoretical proposals to use linear chains and ladders of Josephson junctions to create novel quantum phases and devices, only modest progress has been made experimentally. One major roadblock has been understanding the role of disorder in such systems. We present experimental results that establish the insulating state of linear chains of sub-micron Josephson junctions as Luttinger liquids pinned by random offset charges, providing a one-dimensional implementation of the Bose glass, strongly validating the quantum many-body theory of one-dimensional disordered systems. The ubiquity of such an electronic glass in Josephson-junction chains has important implications for their proposed use as a fundamental current standard, which is based on synchronisation of coherent tunnelling of flux quanta (quantum phase slips). 
\end{abstract}

\maketitle

The combined effects of interaction and disorder in superfluid bosonic condensates can have drastic consequences, leading to the Mott insulator\cite{jaksch-etal-98,greiner-etal-02} and Bose-Anderson glass\cite{fisher-etal-89,giamarchi-schultz-88,giamarchi-book}. The latter is thought to describe helium-4 in porous media, cold atoms in disordered optical potentials, disordered magnetic insulators, and thin superconducting films. The prototypical Bose-Hubbard model without disorder predicts a Beresinskii-Kosterlitz-Thouless quantum phase transition between superfluid and Mott insulator. Experimental implementation using arrays of Josephson junctions has been explored\cite{bradley-doniach-84,geerlings-etal-89,chow-etal-98}, however, the possibility of the insulating glass has not been considered. 

One-dimensional arrays of Josephson junctions are notable for application as a fundamental current standard\cite{zaikin-etal-97,pop-etal-10}, which is based on synchronisation of a `dual' Josephson effect, envisioned to arise from coherent quantum tunnelling of flux quanta, or so-called quantum phase slips\cite{mooij-nazarov-06,guichard-hekking-10,haviland-10,rastelli-etal-13,ergul-etal-13}. Unlike the Mott insulator, the insulating glass is compressible, therefore AC synchronisation of charge may not be possible. Although the presence of offset charge disorder is well-established for small superconducting islands, it has not been sufficiently addressed in regards to dual Josephson effects. 

We have measured critical voltages for a large number of simple chains of sub-micron Josephson junctions with significantly varying energy scales. We observe universal scaling of critical voltage with single-junction Bloch bandwidth. Our measurements reveal a localisation length exponent that steepens with Luttinger parameter, $K$, arising from precursor fluctuations as one approaches the Bose glass-superfluid quantum phase transition. This contrasts with the fixed exponent found for classical pinning of charge density waves\cite{fukuyama-lee-78}, vortex lattices\cite{larkin-ovchinnikov-79} and disordered spin systems\cite{imry-ma-75}, and is in excellent agreement with the quantum theory of one-dimensional disordered bosonic insulators\cite{suzumura-fukuyama-83,giamarchi-schultz-88,giamarchi-book}. Luttinger liquids (LL's) characteristically obey scaling laws with $K$-dependent exponents, thereby we demonstrate a unique signature of pinned Luttinger liquids using insulating Josephson-junction (JJ) chains. 

A Josephson junction array is described by a coupled quantum rotor model, which is equivalent to a long-ranged Bose-Hubbard model with large average number of bosons $ \langle n \rangle $, per site. The Josephson energy $E_\mathrm{J}$ is related to the hopping matrix element $t$ of the Bose-Hubbard model as $ \langle n \rangle  t \rightarrow E_\mathrm{J}$. The on-site energy $U$ of the Bose-Hubbard model is related to the single-junction Cooper-pair charging energy, $E_\mathrm{CP} \equiv \left (2 e\right )^{2}/2 C_\mathrm{J}$, where $C_\mathrm{J}$ is the junction capacitance. In Josephson junction arrays, a third energy scale, $E_\mathrm{0} =\left (2 e\right )^{2}/2 C_\mathrm{0}$, arises from the inevitable capacitive coupling to ground, $C_\mathrm{0}$. For a one-dimensional chain with only nearest-neighbour junction capacitances $C_\mathrm{J}$, and capacitances to ground, the Coulomb interaction $U_{i j}$ decays exponentially as $U_{i j} \simeq \Lambda E_\mathrm{CP} \exp ( -\vert i -j\vert /\Lambda )$, where the screening length $\Lambda $ is given by $\Lambda  =\sqrt{C_\mathrm{J}/C_\mathrm{0}} =\sqrt{E_\mathrm{0}/E_\mathrm{CP}}$. 

In the insulating state of a one-dimensional chain of junctions, it is more convenient to cast the model in terms of continuous quasicharges $\left \{q_{i}\right \}$, where $q_{i} \equiv \pi Q_{i}/2e$ is proportional to the charge $Q_{i}$ that has flown into junction $i$, rather than discrete island charges $\left \{n_{i}\right \}$\cite{quasicharge}.
In this way, an effective Langrangian is obtained\cite{gurarie2004,vogt-etal-15},
\begin{equation}\mathcal{L} =\frac{1}{2 \pi K } \sum _{i}[\frac{\overset{\text{.}}{q}_{i}^{2}}{v} -v \left (q_{i} -q_{i +1}\right )^{2}] -\sum _{i}\epsilon _{0}(q_{i} +f_{i}), \label{sg_lagrang}
\end{equation}
where charge velocity $v =\sqrt{2E_\mathrm{0} E_\mathrm{J}/}\hbar $, Luttinger parameter $K \equiv \pi  \sqrt{E_\mathrm{J}/2 E_\mathrm{0}}$, and the $f_{i}$ describe random offset charges. The energy $E_\mathrm{0}$ is seen to be the elastic energy for small displacements of quasicharge.

Likharev and Zorin\cite{likharev-zorin}
found that the energy levels for a single current-biased junction are given by periodic Bloch energy bands in quasicharge. The lowest energy band $\epsilon _{0}$ is characterised by its Bloch bandwidth $W$. For large $g =E_\mathrm{J}/E_\mathrm{C P}$ , the energy bands become sinusoidal and $\epsilon _{0} = -\frac{W}{2}\cos (2q)$, with
\begin{equation}W = 16 \sqrt{\frac{E_\mathrm{J} E_\mathrm{C P}}{\pi }} (2 g)^{1/4} e^{ -\sqrt{32 g}}, 
\label{bloch_bndwdth}
\end{equation}
so that in this limit, the continuum version of Eqn. (\ref{sg_lagrang})
describes a sine-Gordon model\cite{haviland-delsing-96}.

In this letter we exploit the critical voltage as a probe of localisation (pinning) length $N_\mathrm{L}$, which can be determined using a generalised depinning theory. 
The classical limit of depinning, as applied to JJ chains, has been discussed recently by Vogt \textit{et al.}\cite{vogt-etal-15}.
Under the assumption of maximal offset charge disorder, with the $f_{i}$ distributed independently for each site, the last term in Eqn. (\ref{sg_lagrang}) becomes random, bounded by $ \pm W/2$. The quasicharge is then pinned in a manner analogous to pinning of an elastic charge density wave by random impurities\cite{fukuyama-lee-78}.

Classical pinning of an elastic object by a random potential arises in many contexts, and is related to the study of interface roughness\cite{brazovskii-nattermann-04}. As found in the context of disorderd spin systems\cite{imry-ma-75}, and
pinning of vortex lattices in type II superconductors\cite{larkin-ovchinnikov-79}, one finds a characteristic length $N_\mathrm{L}$ over which the ground state remains ordered. $N_\mathrm{L}$ is set by competition between distortion of the elastic object, which lowers
the total pining energy, but simultaneously increasing the elastic energy. It is found self-consistently that $N_\mathrm{L}$ has a characteristic power law dependence on the range of the pinning distribution, here, $N_\mathrm{L} \propto W^{ -2/3}$. The depinning force is proportional to the elastic energy, $E_\mathrm{0}$, and inversely proportional to $N_\mathrm{L}^{2}$. For chains larger than $N_\mathrm{L}$, the pinning force is simply the critical voltage divided by the number of junctions in the chain, and therefore, $e V_\mathrm{c}/N \propto E_\mathrm{0}^{ -1/3}W^{4/3}$.

One notes from Eqn. (\ref{bloch_bndwdth}) that the leading order
prefactor of $W$ is a constant times the junction plasma frequency, $\hbar  \omega _{p} =\sqrt{2 E_\mathrm{J} E_\mathrm{CP}}$. 
In order to compare chain families of widely varying $\hbar  \omega _{p}$, and chain length $N$, we express the critical voltage and Bloch bandwidth as dimensionless variables, $v \equiv e V_\mathrm{c}/N \hbar  \omega _{p}$ and $w \equiv W/\hbar  \omega _{p}$, so that in the classical limit,
\begin{equation}
v =a w^{4/3}, \label{classsical_depinning}
\end{equation}
with the prefactor, $a =b \left (K/\Lambda \right )^{1/3}$, with $b$ a constant $\mathcal{O} \left (1\right )$.

Recently, voltage-biased Josephson junction arrays have been described using a dual Josephson picture, where the critical voltage arises from coherent quantum phase slips (QPS)\cite{guichard-hekking-10,pop-etal-10,rastelli-etal-13,ergul-etal-13}.
The phase slip rate across each junction in the large $g$ limit is $W/2h$, and under the assumption of independent phase slips across each junction, the critical voltage of a chain would be $V_\mathrm{c} =N \max \vert d \epsilon _{0} (q)/d q\vert (\pi/2e)$, which for large $g$ becomes $\pi  N W/2e$, leading to $v =\pi  w/2$, that is, an exponent of $1$ rather than $\frac{4}{3}$. The simple QPS picture is thus seen to assume rigid quasicharge across the chain, and ignores offset charge disorder. The assumption
of rigid quasicharge is arguably reasonable in the case of an infinite screening length $\Lambda$, or deep in the incompressible Mott insulating state, but questionable in the compressible Bose glass state. 

So far we have only considered the case of classical depinning. When quantum fluctuations are included \cite{suzumura-fukuyama-83, giamarchi-schultz-88,giamarchi-book}, one finds that the localisation length increases with increasing Luttinger constant,
$K$, such that $N_\mathrm{L} \propto w^{ -2/\left (3 -2 K\right )}$, which diverges at the Bose glass-superfluid (BG-SF) transition, $K_\mathrm{c} =3/2$. The critical voltage then scales as
\begin{equation}v =a w^{\alpha },\text{\quad }\alpha  =4/(3 -2 K).
\label{quantum_scaling}
\end{equation} 
The dominant effect of quantum fluctuations of charge, $K \neq 0$, is seen to change the exponent $\alpha $, as the prefactor $a$ is only very weakly dependent on $K$. 


We have experimentally determined the dependence of the critical voltage on chain length $N$, scaled Bloch bandwidth $w$ (varying both plasma frequency $\omega _{p}$ and $g$), and screening length $\Lambda $,  by fabricating and measuring a large ensemble of Al/AlO$_ {x}$/Al single-junction chains. Several families of devices with different plasma frequencies, controlled by the oxide barrier thickness, were initially fabricated on substrates without ground planes (see Figure 1). Within a device family, we vary the junction area $A$ across the family, in order to geometrically tune $g$ (see  SM\cite{supplementary}). 

Our approach is in contrast with previous studies, which have mostly been carried out using SQUID arrays, for which each serial element of the chain consists of two junctions in parallel forming a low inductance loop. The advantage of using such SQUID arrays is that using a single device, the effective $E_\mathrm{J}$ for an element can be tuned \textit{in situ} by applying an external magnetic field. However, this simultaneously changes both $g$ and $\omega _{p}$. Instead we were motivated to examine junction chains that were as simple as possible to fabricate uniformly, not susceptible to disorder arising from unequal SQUID junctions or variations in loop areas, and unaffected by low-frequency flux noise. Furthermore, we desired to keep the plasma frequency as constant as possible for a given family of devices.
\begin{figure}
	\includegraphics[width=8.6cm]{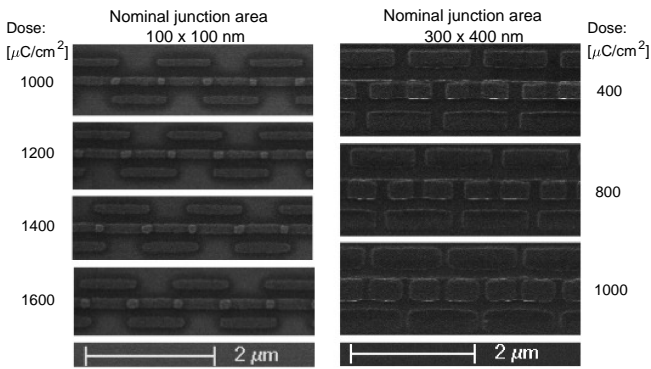}
	\caption{
			SEM~micrograph showing two array families with high (left panel) and low (right panel) plasma frequencies, and nominal junction areas of 100$ \times $100 and 300$ \times $400 nm, respectively. The specific capacitance is 95 fF/$\mu$m\textsuperscript {2} and 54 fF/$\mu$m\textsuperscript {2}  respectively. The precise junction area within a family is modulated by the exposure dose.} 
\end{figure}

For each device, we first obtain an accurate measure of the average junction charging energy, $E_\mathrm{CP}$, from the voltage offset, $V_\mathrm{off}$, of each device found from extrapolating its linear current-voltage characteristic (IVC) from large voltage bias. As noted in \cite{tighe-etal-93,cedergren-etal-15,supplementary}, the experimentally determined charging energy is found as$\text{,}$ $E_\mathrm{CP} =4 e V_\mathrm{off}/N$, and the average Josephson energy $E_\mathrm{J}$ across the chain is found from the normal state conductance using the Abegaokar-Baratoff relation. Note that a given device can be parameterised by either $E_\mathrm{J}$ and $E_\mathrm{C P}$, or alternatively $\hbar  \omega_{p}$ and $g$. 

Next we measure the critical voltage $V_\mathrm{c}$, deep in the subgap region, $V \ll 2 N \Delta/e$ , where $\Delta $ is the superconducting gap. Many of the measured devices have non-hysteretic IVC's in this region. For these devices, the critical voltages are determined to be the voltage at which the current becomes measurably greater than the noise level in the zero-current region. Typically this current is three orders of magnitude or more greater than the zero-current noise level. Some devices with larger $g$ exhibit hysteretic IVC's. For these, the critical voltage is found as the average of the distribution of switching voltages. The switching voltage is defined as the voltage for which the current makes a large jump (maximum $dI/dV$) upon stepping up from zero voltage bias, as illustrated in Fig. 2. The standard deviations of the switching voltages are at most a few percent of the distribution average critical voltage (see SM\cite{supplementary}) for additional details).

 
\begin{figure}
	\includegraphics[width=8.0cm]{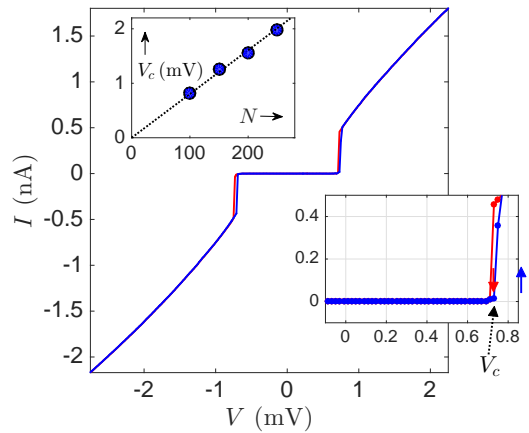}
\caption{(main plot) Experimental determination of the critical voltage for a N = 250, Family 1, device CS3 (see SM\cite{supplementary}).  The blue data is obtained upon stepping up from zero votage, and the red when stepping back down from non-zero current. (lower inset) Close up of the small voltage region where the critical voltage is extracted.  A very small hysteresis region is present in the IVC for this device. The critical voltage is taken to be the average value of the switching voltage, where the latter is defined as the voltage having maximum $dI/dV$ upon stepping up from zero voltage. (upper inset) Linear dependence of $V_c$ on chain length, $N$, for a family of devices where only length has been varied.
}	

\end{figure}

In order to explicitly test the influence of $K\;$on the scaling of critical voltage, we also fabricated and measured a family of devices having a gold ground plane buried under 50 nm of ALD~deposited Al$_ {2}$O$_ {3}$. The presence of the ground plane increases the capacitance to ground $C_\mathrm{0}$, hence lowering $\Lambda $, and therefore increasing $K$ for a given range of $w$, since $K =\pi  \Lambda ^{ -1} \sqrt{g/2}$. Increased $K$ produces a stronger departure from the classical scaling through Eqn.(\ref{quantum_scaling}), as one moves closer to the SF-BG quantum phase transition at $K_\mathrm{c} =3/2$.

In Figure 3, we plot the dimensionless scaled critical voltage, $v$, as a function of the scaled, single-junction Bloch bandwidth, $w =W/\hbar  \omega _{p}$. $W$ has been calculated by numerical diagonalization of the Hamiltonian for a single, current-biased junction, using the experimentally determined values of $E_\mathrm{J}$ and $E_\mathrm{C P}$ for each device. The blue and black dotted lines are the classical expression of the depinning theory, Eqn. (\ref{classsical_depinning})\cite{vogt-etal-15}, for differing $\Lambda$, which are already substantially reduced from the red solid line that arises from a model based on independent, coherent QPS (rigid quasicharge, \text{i.e.} infinite screening length). 

For Figure 3, we performed a least-squares fit to screening length $\Lambda$ and prefactor $b$, using the disordered LL theory, resulting in the solid blue and black lines for devices without and with ground planes, respectively. Note that $K$ is determined from independent measurement of $E_\mathrm{J}$ and $E_\mathrm{CP}$, combined with fit values of the screening length $\Lambda$ (or alternatively $E_\mathrm{0}$). The fitted values of screening length are $\Lambda$\,=\,13.1 for devices without ground planes, and $\Lambda$\,=\,4.0 for the devices with ground plane. The fitted prefactor for devices without ground planes is 11\% larger than the classical value found by Fukuyama and Lee in the context of charge density waves\cite{fukuyama-lee-78}, however, for ground plane devices it is 28\% smaller.  Corrections to the theory arising from slightly non-maximal charge disorder or other microscopic assumptions, would result in a modified prefactor. 

For comparison, we have also included in Figure 3 the disordered LL theory using the Fukuyama-Lee prefactors (dashed blue and black lines), combined with values of $\Lambda $ determined from the observed periodicity in gate voltage, $\Delta U$, of the conductance (see SM\cite{supplementary}). Here one assumes the observed period in the normal state is given by $\Delta U$$C_\mathrm{0}$$ = $$e$, however, additional theory is necessary to understand the experimentally observed periodicities in the transport regime. This approach, which involves no fitting parameters, nevertheless is a very good match to the data, in contrast to the classical result.

For small $w$, $K
\propto \Lambda ^{ -1}  \ln w$, explicitly showing that over the same range of $w$, $K$ is enhanced by the decreased screening length of devices with ground planes, resulting in a stronger departure from the classical result. 

We conclude that the data are clearly inconsistent with the classical depinning theory (slope 4/3), but can be accurately described by the quantum theory, which includes steepening of the localisation length exponent with Luttinger parameter $K$. Note that according to theory, reducing the screening length pushes the classical theory ($K$=0) to higher scaled critical voltages (dotted black line compared to dotted blue), in opposition to the quantum result, which is in excellent agreement with our experiments where we have systematically increased K by engineering the screening length using ground planes.

The universal scaling of the data is furthermore notable as it covers three orders of magnitude in $v$, two in $w$, nearly two in length (N=100-5000), and greater than one in plasma frequency. We therefore identify and demonstrate quantitatively a unique signature of the Bose glass in Josephson-junction chains, as a precursor to Bose-glass-to-superfluid-transition at $K_\mathrm{c}$\,=\,$3/2$, confirming the quantum theory of disordered one-dimensional bosonic insulators based on an interacting Luttinger-liquid picture.

\begin{figure}
	\includegraphics[width=8.6cm]{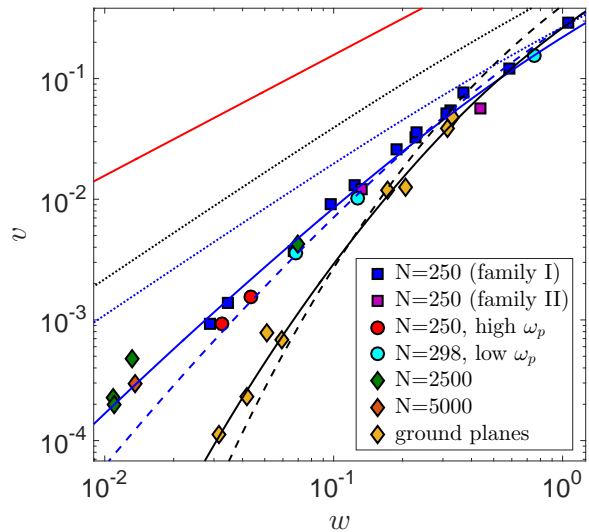}
	\caption{Scaled critical voltage $v =eV_\mathrm{c}/N \hbar  \omega _{p}$, versus scaled Bloch bandwidth $w =W/\hbar  \omega _{p}$. Symbols represent different fabrication `families' distinquised by plasma frequency $\omega_p$, length and presence of ground plane (see SM\cite{supplementary}). The solid red line is theory for independent QPS across each junction (additive Coulomb blockade, or `rigid' quasicharge, and no disorder) and has slope = 1. Solid lines are the quantum theory of a disordered Luttinger mode, Eqn.(\ref{quantum_scaling}), with fitted values of screening length $\Lambda$\,=\,13.1 (blue), and $\Lambda$\,=\,4.0 (black), respectively. These exhibit a $w$-dependent slope, 4/(3\,-\,2$K$), where $K(w)$ is the Luttinger parameter. For small $w$, $K
\propto \Lambda ^{ -1}  \ln w$:  for the same range of $w$, $K$ is enhanced by the decreased screening length of devices with ground planes, resulting in a stronger departure from the classical result. The dotted lines show the classical depinning result, slope = 4/3, for $\Lambda$\,=\,13.1 (dotted blue) and $\Lambda$ = 4.0 (dotted black). Also plotted for comparison are the quantum results for  $\Lambda$\,=\,7.7 (blue dashed), and $\Lambda$\,=\,3.2 (black dashed), screening lengths inferred from the gate-dependent periodicity of $dI/dV$ at large $w$ and biases $V>V_c$, see SM\cite{supplementary}).}
	\end{figure}

Based on a finite size analysis of the zero-bias resistance of SQUID chains, the authors of Ref.\cite{chow-etal-98} concluded that the one-dimensional superfluid-insulator transition occurs at an anomalously low value for the Luttinger parameter\cite{haviland-etal-01,choi-etal-98}. 
In view of our single-junction chain results, which are in remarkable quantitative agreement with theory of the superfluid-Bose glass transition, there appears to be a discrepancy.  We have recently measured SQUID chains which indeed show significantly reduced critical voltages compared to our single-junction chains\cite{duty-unpublished}. Additional experimental work is needed to resolve the matter, which could indicate a non-trivial interplay of flux and charge in SQUID chains.


In the BH model, a sequence of Mott-lobes occur with variation of the chemical potential $\mu$. With disorder in $\mu$, a Bose Glass phase intervenes between Mott insulator and superfluid phases\cite {fisher-etal-89}. For sufficiently strong disorder, the Mott lobes disappear, leaving only the Bose glass. For our devices, the chemical potential is related to gate voltage $U$. We have found no appreciable gate dependence of $V_c$ in any device. This can be understood as a consequence of maximal offset charge disorder, and indicative of the possible ubiquity of the Bose glass phase in insulating Josephson-junction arrays. 

Given that materials used for so-called quantum phase slip devices are significantly disordered, we believe it likely that the Bose glass behaviour we have found in JJ chains may well extend to such superconducting nanowires. In contrast to the rigid Mott insulator, the Bose glass has non-zero compressibility due to low energy rearrangements of domain boundaries. In JJ chains, these are Cooper pairs (or Cooper-pair holes), localised over the pinning length, $N_\mathrm{L}$. Their number and configuration are randomly changed by external voltages. We argue that this could explain the lack of success in achieving sharp current steps under RF or microwave driving, for both junction chains\cite{andersson-etal-00} and superconducting nanowires\cite{lau-etal-01,lehtinen-etal-12}.

\begin{acknowledgments}
	
	This work was supported by the Centre of Excellence for Engineered Quantum Systems, an Australian Research Council Centre of Excellence, CE110001013. Devices were fabricated at the UNSW Node of the Australian National Fabrication Facility. A. Shnirman was supported by the Russian Science Foundation (Grant No. 14-42-00044). N. Vogt was supported by Australian Research Council under the Discovery funding scheme DP140100375, with computational resources were provided by the NCI National Facility systems at the Australian National University through the National Computational Merit Allocation Scheme supported by the Australian Government. A. Shnirman thanks D. G. Polyakov for useful discussions.
\end{acknowledgments}

\bibliographystyle{apsrev4-1} 
\bibliography{xampl}

\begin{thebibliography}{99}
\bibitem {quasicharge}
Quasicharge is well known in the theory of Coulomb blockade and appears naturally in the theories including a phenomenological inductance. It is seen to
be analogous to quasimomentum that arises in the quantum theory of a particle in a periodic potential, see \cite{likharev-zorin}.
\end{thebibliography}

\renewcommand{\thefigure}{S\arabic{figure}}
\setcounter{figure}{0}   
\newpage

\begin{center}
\large  \textbf{Supplemental material to:  ``Insulating Josephson-junction chains as pinned Luttinger liquids"}
\end{center}

\begin{table}
\begin{ruledtabular}
\caption*{N250 Family I}
\begin{tabular}{ C{0.13\linewidth}  C{0.09\linewidth}  C{0.09\linewidth}  C{0.08\linewidth}  C{0.09\linewidth} C{0.09\linewidth} C{0.09\linewidth}  C{0.09\linewidth}   C{0.09\linewidth}  }

Device	& $E_{CP}$ ($\mu$eV) & $E_J$ ($\mu$eV)& $g$ & $\hbar \omega_p$ ($\mu$eV) & $W$ ($\mu$eV) &  $V_{c}$  (mV)& $\log w$ & $\log v$\\ 
\hline 
	AS1   &960  & 24.1 &0.03& 215 &228 &15.6 & 0.03   & -0.54\\	
	AS2    &590  & 40.7& 0.07& 219 &128 & 6.60   & -0.23  & -0.92 \\
	AS6    &437 & 68.4&0.16& 245 &79.1 &  3.30   & -0.49  & -1.27  \\
	AS7    &381  & 87.9 &0.23& 259 &58.8 & 2.10 & -0.64  & -1.49   \\

	BS1   &454 & 60.2&0.13& 234 &86.4  &  4.45   & -0.43  & -1.12  \\
	BS2  &456  & 74.7  &0.16& 261 &81.2 &3.37  & -0.51 & -1.29    \\
	BS3   &403  & 92.0& 0.23& 272 &62.5 & 2.44 & -0.64  & -1.45   \\
	BS4   &376  & 104& 0.28& 279 &52.6 & 1.83   & -0.73 & -1.58   \\
	BS6   &331  & 129&0.39& 293  &36.2 & 0.95  & -0.91 & -1.89    \\
	BS7    &286  & 167&0.58& 309 &20.6 & 0.292  & -1.18  & -2.42  \\

	CS3   &326  & 151 &0.46& 314&30.4 &0.720  & -1.01 & -2.04     \\
	CS5   &259  & 230 &0.89& 345 &9.96 &0.080   & -1.54 & -3.09    \\
	DS1    &272  & 223  &0.82& 348  &12.0 &0.120 & -1.46 & -3.04   \\

\end{tabular}
\end{ruledtabular}

\begin{ruledtabular}
\caption*{N250 Family II (slightly larger junction areas)}
\begin{tabular}{ C{0.13\linewidth}  C{0.09\linewidth}  C{0.09\linewidth}  C{0.08\linewidth}  C{0.09\linewidth} C{0.09\linewidth} C{0.09\linewidth}  C{0.09\linewidth}   C{0.09\linewidth}  }

Device	& $E_{CP}$ ($\mu$eV) & $E_J$ ($\mu$eV)& $g$ & $\hbar \omega_p$ ($\mu$eV) &  $W$ ($\mu$eV)& $V_{c}$  (mV) & $\log w$ & $\log v$\\ 
\hline 
	LAS3  &456  & 48.5 &0.11&210  &91.7 & 3.0 &-0.36 &-1.24    \\
	LAS5   &294  & 109 &0.37& 254 &33.5 &0.77  &-0.88 &-1.91  \\

\end{tabular}
\end{ruledtabular}

\begin{ruledtabular}
\caption*{N=298 Low plasma frequency}
\begin{tabular}{ C{0.13\linewidth}  C{0.09\linewidth}  C{0.09\linewidth}  C{0.08\linewidth}  C{0.09\linewidth} C{0.09\linewidth} C{0.09\linewidth}  C{0.09\linewidth}   C{0.09\linewidth}  }

Device	& $E_{CP}$ ($\mu$eV) & $E_J$ ($\mu$eV)& $g$ & $\hbar \omega_p$ ($\mu$eV) &  $W$ ($\mu$eV)& $V_{c}$  (mV) & $\log w$ & $\log v$\\ 
\hline 
	LPBS2 &56.0  & 21.4 & 0.38 & 49.0  &6.22 &0.15 &-0.90 &-1.99    \\
	LPDS2 &96.5  & 4.43 & 0.05 & 29.2  &22.0 &1.35 &-0.13 &-0.81    \\
	LPDS6 &52.3  & 30.1 & 0.58 & 56.1 &3.84 &0.06   &-1.17 &-2.45 \\

\end{tabular}
\end{ruledtabular}

\begin{ruledtabular}
\caption*{N=250 High plasma frequency}
\begin{tabular}{ C{0.13\linewidth}  C{0.09\linewidth}  C{0.09\linewidth}  C{0.08\linewidth}  C{0.09\linewidth} C{0.09\linewidth} C{0.09\linewidth}  C{0.09\linewidth}   C{0.09\linewidth}  }

Device	& $E_{CP}$ ($\mu$eV) & $E_J$ ($\mu$eV)& $g$ & $\hbar \omega_p$ ($\mu$eV) &  $W$ ($\mu$eV)& $V_{c}$  (mV) & $\log w$ & $\log v$\\ 
\hline 
	HPDS1 &381  & 280 &0.74&462  &20.0 & 0.18 &-1.36 &-2.89   \\
	HPDS4 &331  & 279 &0.84&430  &13.9 & 0.10 &-1.49 &-3.08    \\

\end{tabular}
\end{ruledtabular}

\begin{ruledtabular}
\caption*{N=2500 chains}
\begin{tabular}{ C{0.13\linewidth}  C{0.09\linewidth}  C{0.09\linewidth}  C{0.08\linewidth}  C{0.09\linewidth} C{0.09\linewidth} C{0.09\linewidth}  C{0.09\linewidth}   C{0.09\linewidth}  }

Device	& $E_{CP}$ ($\mu$eV) & $E_J$ ($\mu$eV)& $g$ & $\hbar \omega_p$ ($\mu$eV) &  $W$ ($\mu$eV)& $V_{c}$  (mV) & $\log w$ & $\log v$\\ 
\hline 
FD2500  &266  & 151  &0.57 &283  &19.8 & 3.04 &-1.15 &-2.37  \\
FC2500 &243  & 295  &1.21 &379  &4.98 & 0.45 &-1.88 &-3.32     \\
FB2500 &213 & 276  &1.30 &343  &3.76  &0.17 &-1.96 &-3.70    \\
FA2500&206  & 268 &1.30 &332  &3.60  &0.19 &-1.97 &-3.64   \\

\end{tabular}
\end{ruledtabular}

\begin{ruledtabular}
\caption*{N=5000 chains}
\begin{tabular}{ C{0.13\linewidth}  C{0.09\linewidth}  C{0.09\linewidth}  C{0.08\linewidth}  C{0.09\linewidth} C{0.09\linewidth} C{0.09\linewidth}  C{0.09\linewidth}   C{0.09\linewidth}  }

Device	& $E_{CP}$ ($\mu$eV) & $E_J$ ($\mu$eV)& $g$ & $\hbar \omega_p$ ($\mu$eV) &  $W$ ($\mu$eV)& $V_{c}$  (mV) & $\log w$ & $\log v$\\ 
\hline 
C1A8 &229  & 274 & 1.20 &354 & 4.85 & 0.52& -1.86 & -3.53  \\

\end{tabular}
\end{ruledtabular}

\begin{ruledtabular}
\caption*{N=100-250  Length dependence}
\begin{tabular}{ C{0.13\linewidth}  C{0.09\linewidth}  C{0.09\linewidth}  C{0.08\linewidth}  C{0.09\linewidth} C{0.09\linewidth} C{0.09\linewidth}  C{0.09\linewidth}   C{0.09\linewidth}  }
Device	& $E_{CP}$ ($\mu$eV) & $E_J$ ($\mu$eV)& $g$ & $\hbar \omega_p$ ($\mu$eV) &  $W$ ($\mu$eV)& $V_{c}$  (mV) & $\log w$ & $\log v$\\ 
\hline 
	N=100 &432  & 76.4 &0.18 &257  &74.9 & 0.81 &-0.54 &-1.50   \\
	N=150 &448  & 75.5  &0.17 &260  &79.0 &1.25 &-0.52 &-1.49   \\
	N=200 &412  & 77.3  &0.19 &252  &69.8 &1.56 &-0.56 &-1.51 \\
	N=250 &402  & 75.3 &0.19 &246  &68.1  &1.98 &-0.56  &-1.49  \\

\end{tabular}
\end{ruledtabular}

\end{table}

\begin{table}

\begin{ruledtabular}
\caption*{N=250 with ground plane}
\begin{tabular}{ C{0.13\linewidth}  C{0.09\linewidth}  C{0.09\linewidth}  C{0.08\linewidth}  C{0.09\linewidth} C{0.09\linewidth} C{0.09\linewidth}  C{0.09\linewidth}   C{0.09\linewidth}  }

Device	& $E_{CP}$ ($\mu$eV) & $E_J$ ($\mu$eV)& $g$ & $\hbar \omega_p$ ($\mu$eV) &  $W$ ($\mu$eV)& $V_{c}$  (mV) & $\log w$ & $\log v$\\ 
\hline 
AS2 &406  & 61.8  &0.15&224& 74.1  & 2.65 &-0.48 & -1.33 \\
AS1 &397  & 64.1 &0.16&226& 71.1 & 2.20&-0.50& -1.41\\

\end{tabular}
\end{ruledtabular}

\begin{ruledtabular}
\caption*{N=2500 with ground plane}
\begin{tabular}{ C{0.13\linewidth}  C{0.09\linewidth}  C{0.09\linewidth}  C{0.08\linewidth}  C{0.09\linewidth} C{0.09\linewidth} C{0.09\linewidth}  C{0.09\linewidth}   C{0.09\linewidth}  }

Device	& $E_{CP}$ ($\mu$eV) & $E_J$ ($\mu$eV)& $g$ & $\hbar \omega_p$ ($\mu$eV) &  $W$ ($\mu$eV)& $V_{c}$  (mV) & $\log w$ & $\log v$\\ 
\hline 
C2BA1  &248  & 154  &0.62&275& 16.7  & 0.47& -1.22& -3.17 \\
C2BA2  &226  & 153  &0.68&264& 13.3  & 0.51& -1.30& -3.11 \\
C2XA2  &315  & 94.6  &0.30&244&41.8 & 7.20 & -0.77 & -1.93 \\

C2XA1  &373 & 95.0  &0.25&266& 54.7 & 8.30 & -0.69 & -1.90 \\
C4CA1  &213  & 159  &0.75&260&10.9  & 0.15 & -1.38& -3.64\\
C4CA2  &192  & 164  &0.85&251&7.83  & 0.07& -1.50& -3.96 \\

\end{tabular}
\end{ruledtabular}

\end{table}

\section{Samples}
Al/AlO$_{x}$/Al junction chains were fabricated by standard e-beam lithography followed by two angle evaporation of aluminum with \textit{in situ} oxidation between the two evaporation steps. Each film is 30 nm thick. Devices were fabricated either on silicon substrates (n-doped) with approximately 300 nm thermally grown SiO\textsubscript {2} on top, or, in order to achieve larger capacitance  to ground, on silicon substrates with a 30 nm gold film covered by 50 nm Atomic Layer Deposition (ALD) grown Al$_ {2}$O$_{3}$. A total of 31 devices without ground planes were measured, in addition to 6 devices with buried ground planes. 

The experimentally determined and calculated parameters for the devices, organized by device family, are listed in tables below. The junction Josephson energy, $E_\mathrm{J}$, and Cooper-pair charging energy, $E_\mathrm{CP}$, were determined from large scale current-voltage characteristics (IVC's) in the Ohmic regime, voltage bias $eV \gg 2NE_{C}$, where $E_{C}\equiv e^2/2C_J$ is the single-electron, junction charging energy, as described below. and the average Josephson energy $E_\mathrm{J}$ across the chain is found from the normal state conductance using the Abegaokar-Baratoff relation. A given device can be parameterised by either $E_\mathrm{J}$ and $E_\mathrm{C P}$, or alternatively $\hbar  \omega_{p}=\sqrt{2E_J E_{CP}}$ and $g=E_J/E_{CP}$. The Bloch bandwidth, $W$, was numerically determined by diagonalization of the single-junction Hamiltonian. Critical voltage $V_c$ was experimentally determined as described below, and finally, scaled quantities $v=eV_c/\hbar\omega_p$ and $w=W/\hbar\omega_p$ were calculated. 

\section{Josephson energy}

$E_\mathrm{J}$ is determined from the linear conductance at large voltage bias, $eV \gg 2N\Delta$, using the Ambegaokar-Baratoff relation $E_\mathrm{J} =\frac{\Delta }{2} \frac{R_\mathrm{Q}}{R_\mathrm{J}}$, where $R_\mathrm{Q}$ is the superconducting resistance quantum, and $R_\mathrm{J}$ the inverse of the large-bias linear conductance $G_{\Omega }$ of the array divided by $N$. Here $\Delta $ is taken to be 210 $\mu$eV, as found in \cite{cedergren-etal-15}. Figure S1 (bottom) shows the conductance of N=250, Family I device AS1, which was found to have a total resistance $1/G_{\Omega }$\,=\,7.02 M$\Omega$, giving $R_\mathrm{J}$\,=\,28.1 k$\Omega$, and therefore $E_\mathrm{J}$\,=\,24.1 $\mu$eV.

\section{Charging energy}
$E_\mathrm{CP}$ is extracted from the voltage offset, $V_\mathrm{off}$, of each device, found from extrapolating its linear IVC from large voltage bias. This is a standard procedure which has been used previously by many groups, and is described in \cite{cedergren-etal-15, tighe-etal-93, kuzmin-etal-92, delsing-92}. The experimentally determined charging energy is found as, $E_\mathrm{CP} = 4 E_{C}=4 e V_\mathrm{off}/N$. One must take care that $V_\mathrm{off}$ is extrapolated from sufficiently high voltages above the Coulomb blockade where the conductance $d I/d V$\ is measured to be constant, in order to get an unbiased estimate of $E_\mathrm{CP}$. This is illustrated in Fig. S1 for N=250, Family I device AS1, finding $V_{\rm off}$\,=\,60.0 mV for this device, giving $E_C=eV_{\rm off}/N$\,=\,240 $\mu$eV, or $E_{CP}=4E_C$\,=\,960 $\mu$eV.

\section{Tuning of Josephson-to-Coulomb energy ratio \lowercase{$g$}}
Since chains of simple junctions, in contrast to SQUID chains, do not offer \textit{in situ} control over $g =E_\mathrm{J}/E_\mathrm{CP}$, it has been necessary to fabricate large numbers of arrays with a slight variation in control parameter $g$ between devices. If the oxide barrier is kept the same, assuming a uniform barrier, $E_\mathrm{J}$ should grow linearly with the number of transverse channels and hence with junction area $A$. On the other hand, assuming a parallel plate capacitance for the junction, the charging energy should grow as the inverse of $A$. As a result $g$ scales as $A^{2}$. At the same time the plasma frequency $\omega _{p} =\sqrt{2E_\mathrm{J}E_\mathrm{CP}}$ stays more or less constant. The area within a family of devices, i.e. with the same plasma frequency, has been varied between devices by varying the exposure dose during e-beam exposure and thereby producing proximity effects leading to a larger area for larger doses. This method is particularly effective for small junction sizes. In order to change the plasma frequency, different families of devices with different oxidation conditions have been fabricated. Between such families the nominal area may have to be changed significantly to keep $g$ in the desired range, as was shown in Fig. 1 of the main text.

\section{\textbf{Critical voltage}}
Following the characterisation of the Coulomb and Josephson energy scales for each device, we measure the critical voltage $V_\mathrm{c}$, at biases deep in the subgap region, $eV \ll 2N\Delta $. Devices with low values of $g$ typically have non-hysteretic IVC's in this region.
For these devices, the critical voltages are determined to be the voltage at which the current becomes measurably greater than the noise level in the zero-current region. 

As an example, the subgap IVC of ground plane device AS1 is shown In Figure S2. For this measurement, the zero-current noise level is less than 60 fA, and the device switches to 20 pA. For devices in this category, we integrate the current measurement sufficiently to reduce the zero-current noise level, such that the current switches to greater than two, and typically three, orders of magnitude greater than the subgap noise level for determination of the critical voltage. The limiting current resolution of our setup was determined to be less than 1 fA for long integration times. 

For several of these devices we unsuccessfully looked for evidence for a non-zero current in the voltage gap region due to charge `creep'. However, we did not observed such currents above our limiting resolution, at base temperatures of less than 20 mK. Future experiments could aim to measure thermally activated creep in the temperature window between base temperature and the parity temperature $T^*\simeq$ 300mK\cite{cedergren-etal-15}.

At larger values of $g$, some devices exhibit hysteretic IVC's. For these devices, the critical voltage is found as the average of the distribution of switching voltages. The switching voltage is defined as the voltage for which the current makes a large jump (maximum $dI/dV$) upon stepping up from zero voltage bias. This is illustrated in Figure S3 for Family I device SC5. The inset of Figure S3 shows a histogram of the switching voltages giving $V_c=\left<V_{\rm sw}\right>$=80 $\mu$V, with a standard deviation of 6 $\mu$V.The standard deviations of the switching voltages are typically only a few percent of the distribution average critical voltage. In principle one could represent this as an error bar on scaling plots such as Figure 3 in the main text. However, for our measured devices, it would be unresolvable on such a log-log plot.

To reach higher values of $g$,  we needed to increase the length of the array. For example,  Fig. S4 (left panel) shows two arrays with N = 250 and $g$ = 0.89 (blue data), and $g$ = 1.3 (red data). The critical voltage in the N=250 with larger $g$ is not resolvable, but by measuring an array with the same $g$, but having a length an order of magnitude larger, N = 2500, one observes a well-defined critical voltage.

\section{\textbf{Screening length inferred from gate-periodicity}}

In the experiments reported by Tighe \textit{et al.}\cite{tighe-etal-93}, which were carried out on short ($N$=70) devices very deep in the insulating state ($g\ll1$), the screening length $\Lambda$ for one device having $g<0.01$ was inferred from gate periodicity of the critical voltage. These authors found the same period in gate voltage, $U$, for both the superconducting state (zero magnetic field), and in the normal state obtained when a large magnetic field was used to completely suppress superconductivity. It was therefore assumed by Tighe \textit{et al.} that the gate periodicity, $\Delta U$, was equal to $e/C_0$, allowing a determination of $C_0$, which combined with $C_J$ from the estimate of charging energy discussed above, allows an estimate of the screening length, $\Lambda=\sqrt{C_J/C_0}$. 

For our devices, the gate voltage is given by $U =(V^{ -} +V^{ +})/2$ where $V^{ +}$ and $V^{ -}$are the voltages on the high and low side of the device, respectively. The measured critical voltages for all devices here showed no appreciable gate dependence in the superconducting state.

Shallow gate dependence for the conductance, $dI/dV$, at voltage biases above $V_c$ was observed, however, for 5 devices without ground planes, and 3 with ground planes, all with relatively low values of $g$. Figure S5 shows the gate dependence of $dI/dV$ for Family I device AS2, which is reminiscent of the stability diagram of a single-electron transistor. Devices without ground planes showed the same period, $\Delta U$, in both the superconducting state, and the normal state. Devices with ground planes, however, showed a period doubling of $\Delta U$ in the superconducting state, compared to the normal state, which can be considered as a magnetic-field induced parity effect.

As in Tighe \textit{et al.}\cite{tighe-etal-93}, we can extract estimates of the screening length, by assuming the gate periodicity satisfies, $\Delta U=e/C_\mathrm{0}$, where $\Delta U$ is the normal state  period. Using this method, the inferred screening length is found to be fairly constant within each group of devices. Based on these measurements, using the statistics within each of the two groups, we infer $\Lambda=7.7(0.7)$ for devices without ground planes, and $\Lambda=3.2(0.3)$ for devices with ground planes, where the numbers in parentheses refer to the standard deviation within each group. One can question, however, the validity of such estimates of $\Lambda$, since one uses data in the transport regime, at voltage bias considerably higher than $V_c$. New theory is necessary to examine the validity of such an approach.

\section{Microwave filtering}

 Sufficient microwave filtering is essential in order to resolve critical voltages, especially for devices at larger values of $g=E_J/E_{CP}$. Microwave filtering of DC lines was achieved using three meters of Thermocoax(R) running from room temperature to base temperature, with thermal anchoring at each stage, in combination with LC filters on the circuit board. This setup has a measured cut-off frequency of approximately 1MHz\cite{cedergren-etal-15}. In addition, low pass filters were placed at the room temperature connections. Fig. S4 shows the effect of replacing twisted-pair looms with Thermocoax lines for the voltage bias. This facilitated observation of insulating behavior at larger values of $g$.

\captionsetup{justification=raggedright,singlelinecheck=false}

\begin{figure}
	\includegraphics[ width=8.6cm,]{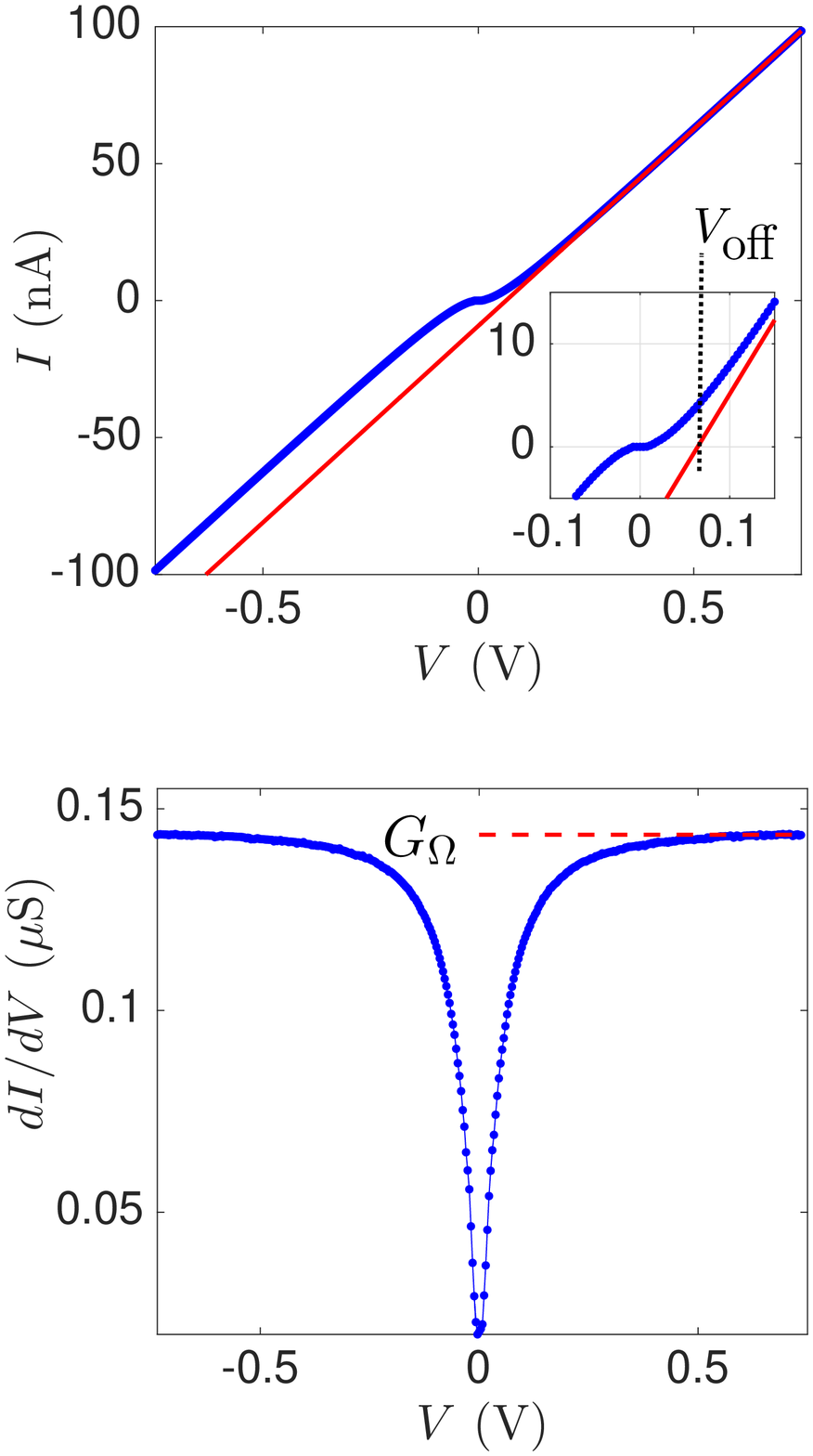} 
	\caption{Large scale IVC data from N\,=\,250 Family I, device AS1. The offset voltage, $V_\mathrm{off}$, (top) is extracted by interpolation from a large $V_\mathrm{bias}$ where $dI/dV$ is flat (bottom). The top inset shows the experimentally determined $V_{\rm off}$\,=\,60.0 mV for this device, giving $E_C=eV_{\rm off}/N$\,=\,240 $\mu$eV, or $E_{CP}=4E_C$\,=\,960 $\mu$eV. The bottom plot shows the conductance, having total resistance $1/G_{\Omega }$\,=\,7.02 M$\Omega$, giving $R_\mathrm{J}$\,=\,28.1 k$\Omega$, and therefore $E_\mathrm{J}$\,=\,24.1 $\mu$eV. }
	
\end{figure}

\begin{figure}
\includegraphics[ width=8.6cm,]{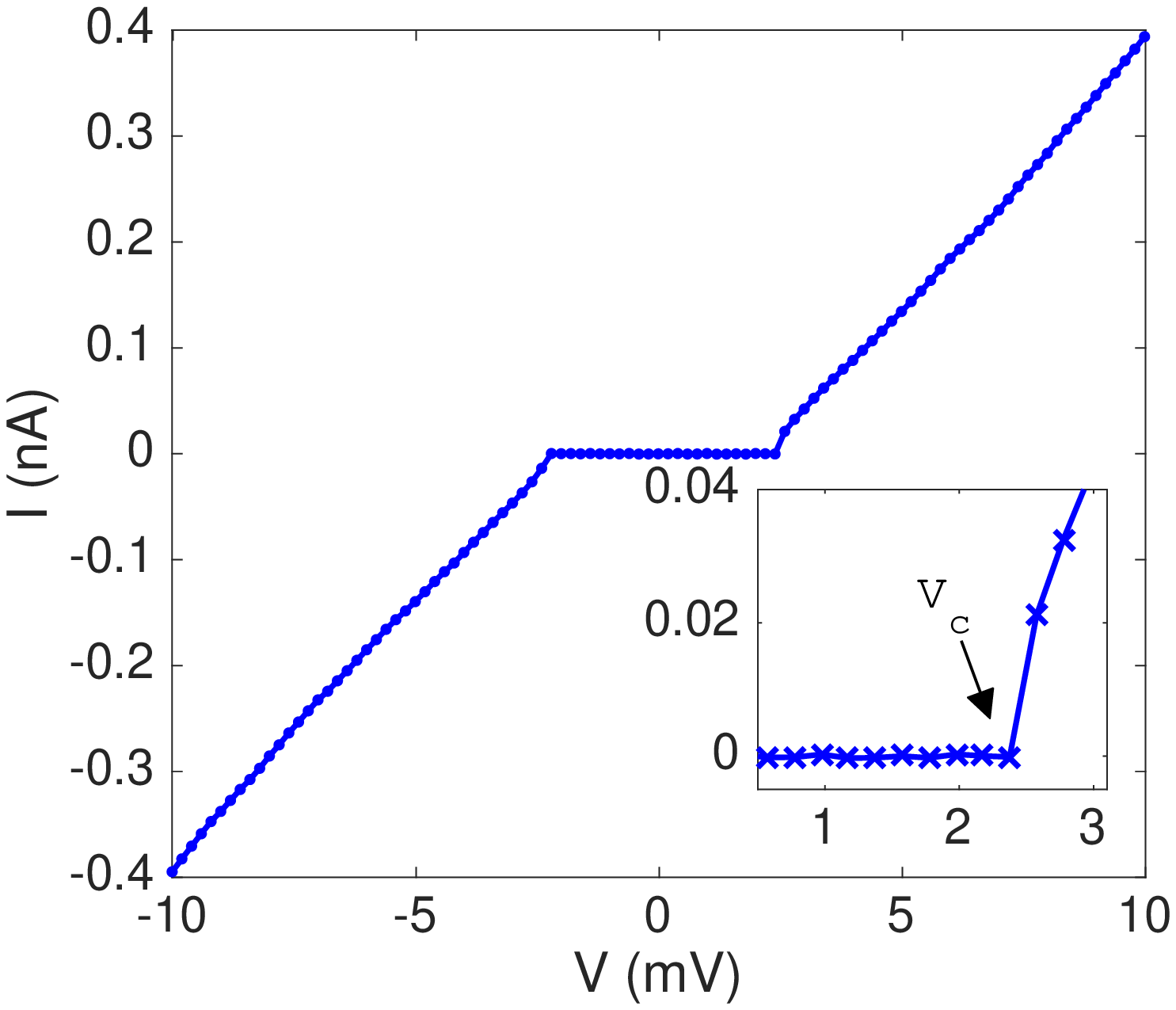}
\caption{IVC for the N=250, ground plane device AS1. The inset shows a zoom into the region of the critical voltage. The zero-current noise level in the voltage gap is less than 60  fA, and the device switches reproducibly to 18 pA at $V_c$\,=\,2.20 mV, giving a well-defined  critical voltage.}
\end{figure}

\begin{figure} 
\includegraphics[ width=8.6cm,]{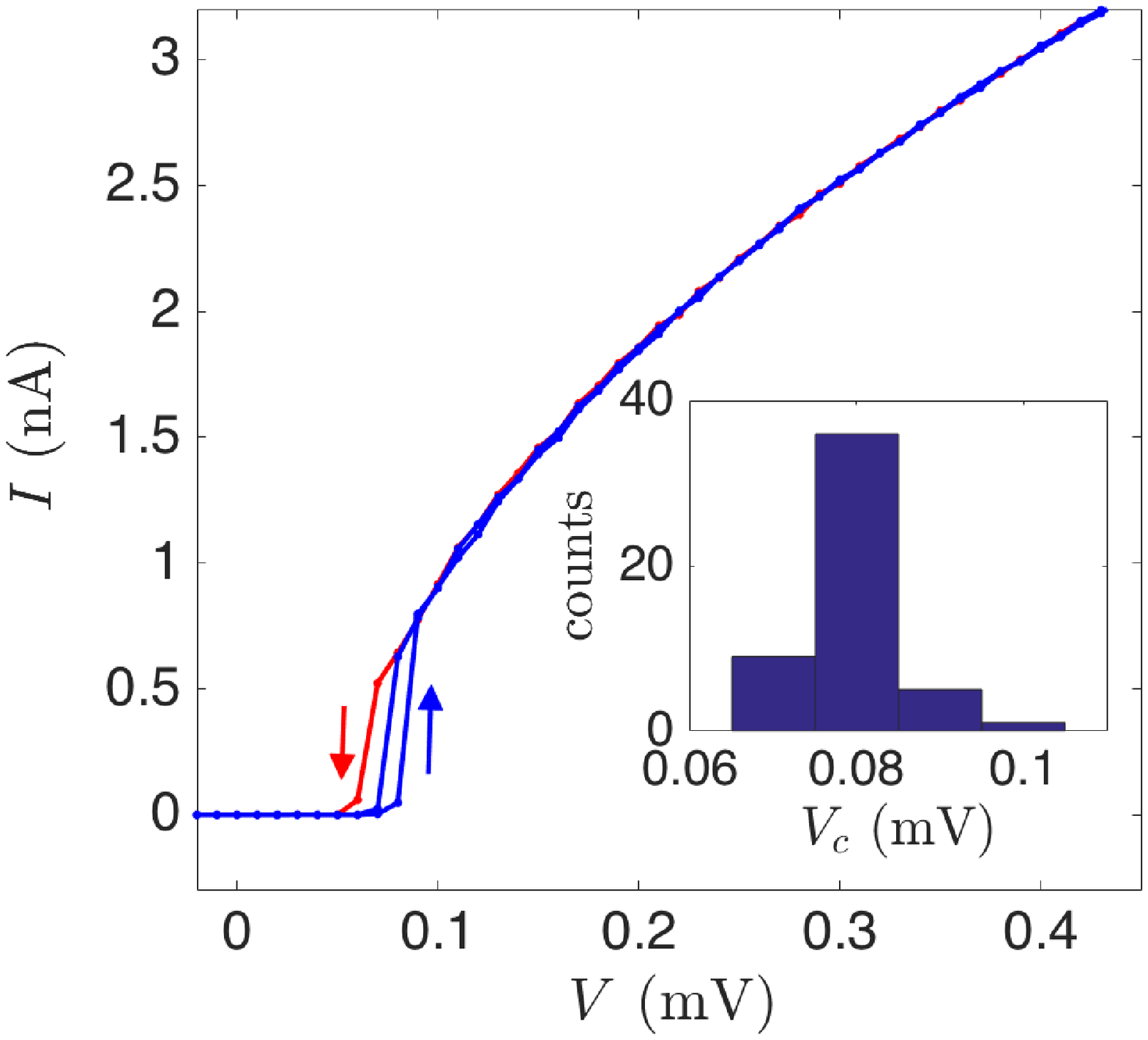}
\caption{Measurements of switching voltage for N=250 Family I, device SC5. The switching voltage $V_{\rm sw}$ is defined as the voltage for which the current makes a large jump (maximum $dI/dV$) upon stepping up from zero voltage bias, as shown for two repetitions (blue) in the main panel. The inset shows a histogram of measurements of $V_{\rm sw}$, which give $V_c=\left<V_{\rm sw}\right>$=80 $\mu$V, and a standard deviation, $\sigma_{\rm sw}$ = 6 $\mu$V.}
\end{figure}

\begin{figure}
\includegraphics[ width=8.6cm,]{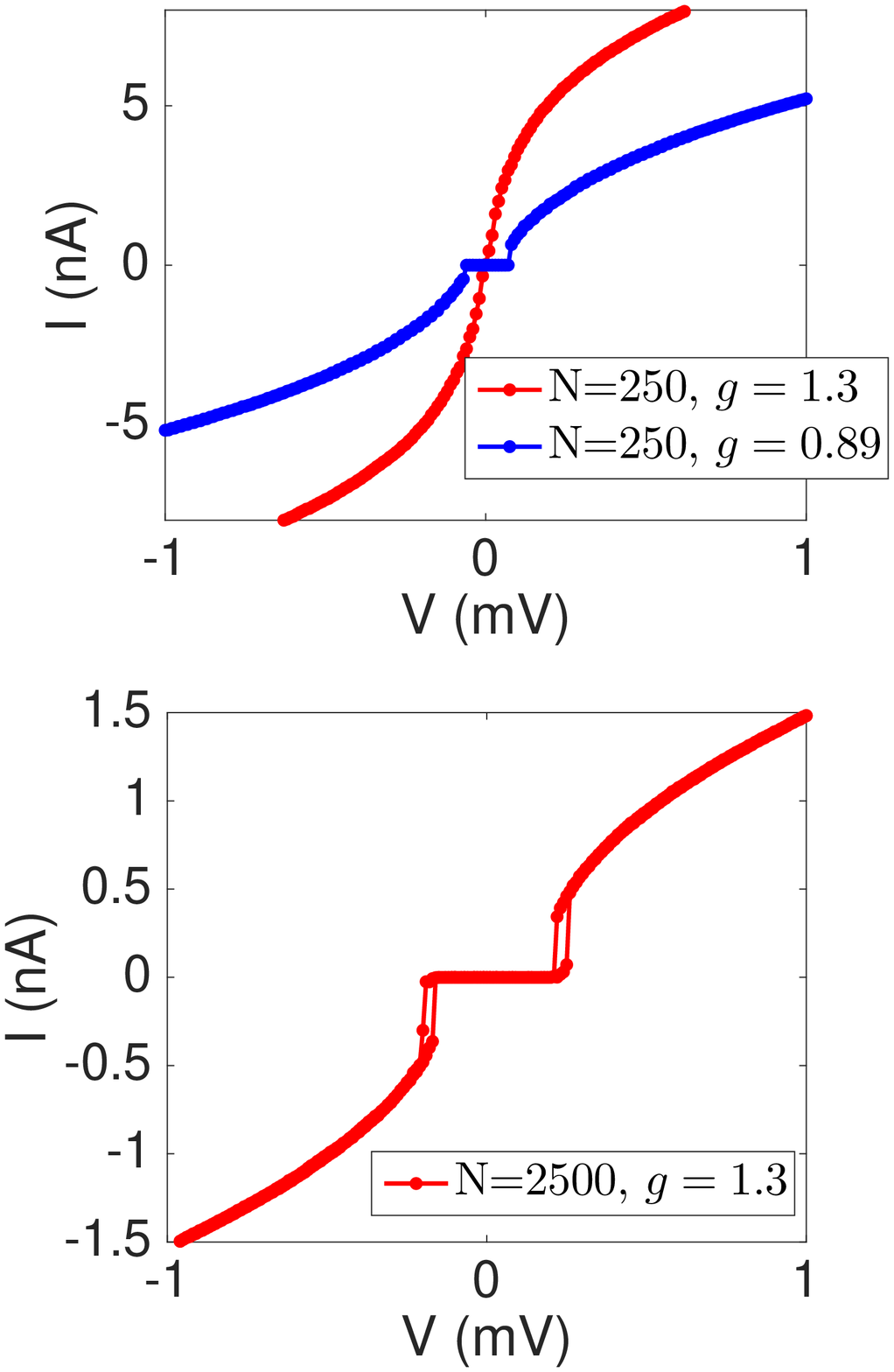}
\caption{IV characteristics of two chains, 250 junctions long, with $g$ = 0.8 and $g$ = 1.3 respectively (left panel). The critical voltage in the N = 250 chain with $g$ = 1.3 can not be resolved, however,  a 10 times longer array with identical $g$ (right panel) shows a well-defined critical voltage.}
\end{figure}

\begin{figure}
\includegraphics[ width=8.6cm]{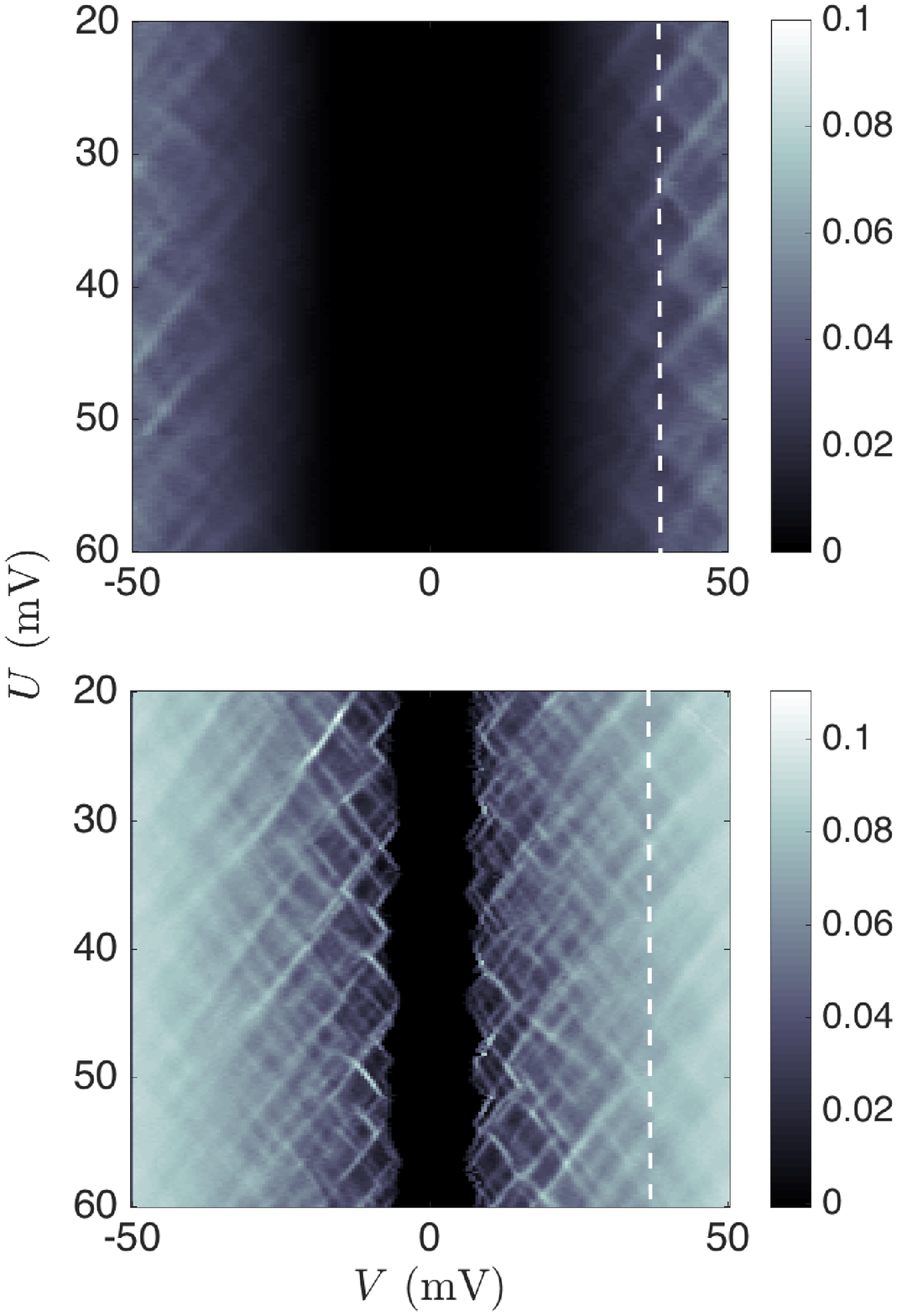} 
\caption{Gate modulation of $dI/dV$ for an array (top), in the superconducting regime (zero applied magnetic field), and (bottom), the normal state at a large applied magnetic field. The dashed lines show the bias voltages for which the periodicity, $\Delta U$, has been extracted.}
 
\end{figure}

\begin{figure}
\includegraphics[ width=8.6cm]{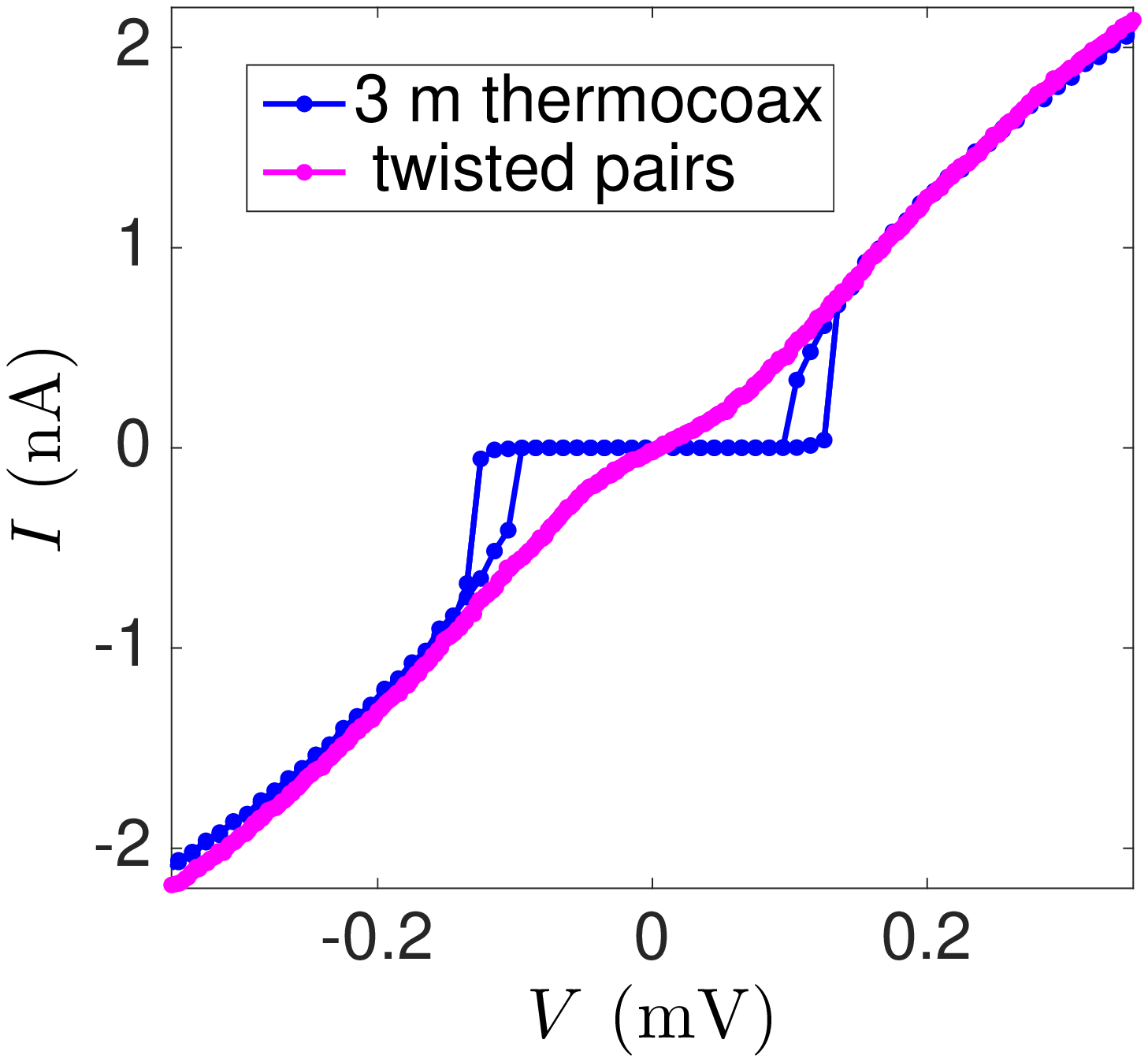}
\caption{The same device, $g$ = 0.82, measured with 3 m of thermocoax microwave filtering running from room temperature to base temperature, compared to using twisted pairs.} 
\end{figure}

\end{document}